Cite as:

**IEEE**

M. Farhadi and M. Abapour, "Three-Switch Three-Phase Inverter With Improved DC Voltage Utilization," in *IEEE Transactions on Industrial Electronics*, vol. 66, no. 1, pp. 14-24, Jan. 2019.

**Plain Text**

M. Farhadi and M. Abapour, "Three-Switch Three-Phase Inverter With Improved DC Voltage Utilization," in *IEEE Transactions on Industrial Electronics*, vol. 66, no. 1, pp. 14-24, Jan. 2019.

doi: 10.1109/TIE.2018.2829680
URL: http://ieeexplore.ieee.org/stamp/stamp.jsp?tp=&arnumber=8345776&isnumber=8453958

**BibTeX**

@ARTICLE{8345776,

author={M. {Farhadi} and M. {Abapour}},

journal={IEEE Transactions on Industrial Electronics},

title={Three-Switch Three-Phase Inverter With Improved DC Voltage Utilization},

year={2019},

volume={66},

number={1},

pages={14-24},

keywords={DC-AC power convertors;invertors;predictive control;switching convertors;transient response;voltage control;gate drive;control circuit components;output filter;voltage utilization factor;four-switch three-phase inverter;six-switch inverter;sine pulsewidth modulation;semiconductor switches;dc-ac converter;model predictive controller;power switch;three-phase universal inverter;Inverters;Capacitors;Inductors;Switches;Topology;Standards;Predictive control;DC–AC power converters;model predictive control (MPC);pure sinusoidal output voltage;three phase},

doi={10.1109/TIE.2018.2829680},

ISSN={},

month={Jan},}

# Three-Switch Three-Phase Inverter with Improved DC Voltage Utilization

Masoud Farhadi, and Mehdi Abapour

*Abstract*—This paper proposes a new DC–AC converter with a less number of semiconductor switches and therefore gate drive and control circuit components. A salient feature of this inverter is its pure sinusoidal line-to-line voltages with no need for output filter. Also, the proposed inverter improves the voltage utilization factor of the input dc supply compared to four-switch three-phase inverter (first best topology in the literature from the number of switches point of view) and standard six-switch inverter with sine PWM. The proposed inverter is capable of operating with a wide range of output voltages from zero to the full value of the dc input voltage by appropriately altering instantaneous duty-cycle. For the first time to our knowledge, in this paper, a three-phase universal inverter with only three power-switch has been proposed. For the purpose of providing a good compromise between fast transient response and stability, a model predictive controller is proposed. All the design expressions have been derived. A comparison with other DC–AC converters is given to show the merits of the proposed converter. Finally, satisfactory circuit operation is confirmed by experimental results from a laboratory prototype.

*Index Terms*—DC–AC power converters, three phase, pure sinusoidal output voltage, model predictive control (MPC).

## I. INTRODUCTION

NUMEROUS DC–AC converters and wide variety of control strategies have been proposed in the recent years. In spite of all the advantages introduced by these converters, the presence of more number of semiconductor switches reduces efficiency and reliability [1]-[3], in addition increases the cost of system. Therefore, past decade has witnessed an increasingly growing research interest in power electronics for reduced switch count converters with as many desirable features of previous converters as possible.

Various reduced switch count topologies have been proposed in the literature for multilevel DC-AC converters. These inverters could be classified into two major categories: 1) inverters without H-bridge; 2) inverters with H-bridge. The first category could be also divided into three categories: a) cascaded bipolar-switched-cells-based multilevel inverters [4], b) packed-U cell inverters [5], [6], and c) T-type inverters [7], [8]. The second category could be also divided into six categories: a) multilevel module-based multilevel inverters [9], b) reversing voltage-based multilevel inverters [10], c) two-switch enabled level-generation based multilevel inverters [11], d) series-connected switched sources-based multilevel inverters [12], e) switched series/parallel sources-based multilevel inverters [13], and f) cascaded half-bridge-based multilevel DC-Link inverters [14]. These inverters have their own advantages and disadvantages from the point of view of application requirements.

However, only some reduced switch count topologies have been proposed for two level three-phase inverter [15]-[17]. Four-switch three-phase converter [18] is the pioneering effort initiated with the aim of decreasing the number of power switches by DC-bus midpoint connection. The four-switch three-phase inverter is widely investigated in several applications after its introduction in 1984 by Broeck and Wyk [18]. Beyond the reduced number of switches by this topology, the fault-tolerant capability [19], reduced number of interface circuits, and more reliability [20]-[22] make it attractive for many critical applications. However, the four-switch three-phase inverter is known to have several disadvantages compared to standard six-switch three-phase inverter. First, the voltage utilization factor is halved compared to the standard three-phase inverter. This restricts the load and speed range of machine, where the input voltage is lower than the rated voltage of motor (such as PV and battery). Second, the imbalanced input ac currents, and the circulation of one-phase current through the split dc-link voltage sources are not suppressed. As a result, the capacitor center tap voltage fluctuation is inevitable; therefore, its application is limited. Therefore, considerable research efforts have been made to improve the performance of the four-switch three-phase inverter. Regarding the carrier based and space vector modulation schemes, considerable researches have been focused on the suppression of the capacitor center tap voltage fluctuation [23]–[25]. Space vector modulation schemes could be classified into three major categories: 1) the equivalent zero vector by two large-amplitude opposite vectors [26], [27]; 2) the equivalent zero vector by two small-amplitude opposite vectors [28], [29]; and 3) the nearest three vectors SVM method [30].

In [31], the authors present a three-switch three-phase inverter that only applicable for induction motor drives. In this topology, the non-torque producing zero sequence current is significantly increased, which necessitates oversized DC-link capacitors. Also, this inverter is only used in situations where the DC-link midpoint and the motor neutral-point can both be accessed.

To the authors' knowledge, this is the first time a three-phase universal inverter with only three power switch has been proposed (see Fig. 1). The proposed converter has better voltage utilization factor of the input dc supply compared to both four-switch three-phase inverter and standard six-switch three-phase inverter with SPWM controller. Also, the proposed inverter does not suffer from the problems of circulation of one-phase current through the split dc-link capacitors. In this topology, three-phase output voltages are pure sinusoidal waves and total harmonic distortion is minimized.

## II. PROPOSED TOPOLOGY AND OPERATION MODES

Fig. 1 shows the proposed three-switch three-phase inverter. This inverter contains one dc power supply, two intermediate capacitors ($C_1$, $C_2$), two output capacitors ($C_3$, $C_4$), three inductors, and three active power switches ($Q_x$, $Q_1$, $Q_2$). As shown in Fig. 1, the basic idea behind control strategy is to produce a set of dc-biased sinusoidal voltages (second phase and third phase) with a phase difference of 60° and connect the first phase to the input dc supply to maintain balanced line-to-line voltages.

Fig. 2 shows phasor diagrams of proposed inverter and standard six-switch inverter. The reference voltages of proposed inverter are considered as follows:

$$V_{AN} = V_{dc} \quad (1)$$

$$V_{BN} = V_{dc} + V_m \sin(\omega t) \quad (2)$$

$$V_{CN} = V_{dc} + V_m \sin(\omega t - 60°) \quad (3)$$

Under this assumption, a balanced set of line-to-line voltages is obtained as follows:

$$V_{AB} = V_A - V_B = -V_m \sin(\omega t) = V_m \sin(\omega t + \pi) \quad (4)$$

$$V_{BC} = V_m \sin(\omega t + \frac{\pi}{3}) \quad (5)$$

$$V_{CA} = V_m \sin(\omega t - \frac{\pi}{3}) \quad (6)$$

It is important to note that the lower and upper bounds for $V_m$ are imperative to guarantee the proposed inverter to work in balanced modes. As a result, the constraint for $V_m$ (peak of line-to-line output voltage) is set as follows:

$$0 \leq V_m \leq V_{dc} \quad (7)$$

In practical situation, the lower bound is selected to be slightly higher than zero. The instantaneous load currents are obtained as follows:

$$i_A = i_m \sin(\omega t + \frac{5\pi}{6} - \varphi) \quad (8)$$

$$i_B = i_m \sin(\omega t + \frac{\pi}{6} - \varphi) \quad (9)$$

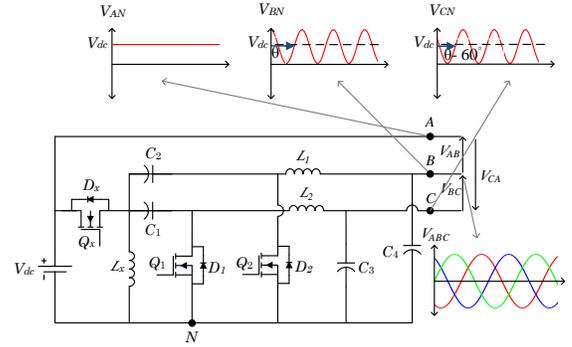

Fig. 1. Proposed three-switch three-phase inverter.

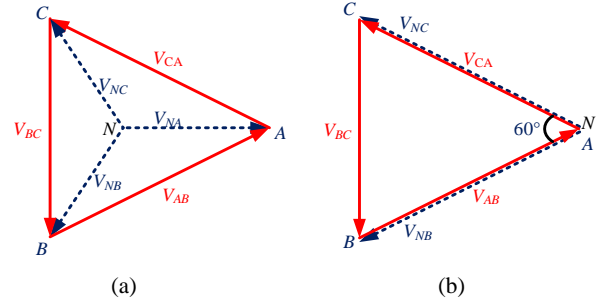

Fig. 2. Phasor diagrams. (a) Standard six-switch inverter. (b) Proposed inverter.

$$i_C = i_m \sin(\omega t - \frac{\pi}{2} - \varphi) \quad (10)$$

There are three power switches with two states in the proposed converter. Therefore, eight ($2^3$) possible switching states could be obtained. There is no prohibited state that produce either a short circuit or a situation in which the switches would have to absorb the inductive energy instantly.

Fig. 3 shows the all possible switching states of proposed inverter where "1" means is the corresponding switch turned ON. The different modes of inverter in state-space form can be expressed in terms of switching signals, capacitor voltages, inductor currents and dc-side voltage, as (11)-(13). The general behavior of different modes is described as follows:

*Mode 1* $(Q_x, Q_1, Q_2) = (1,0,0)$: The input supply provides energy to both loads (phase B and phase C) as well as to the inductors ($L_x, L_1, L_2$). In this case, $V_B$ and $V_C$ will be increased.

*Mode 2* $(Q_x, Q_1, Q_2) = (1,1,0)$: In this mode, the input supply provides energy to phase B as well as to the inductors ($L_x, L_2$) and some of output capacitor ($C_3$) energy will be transferred to the inductor $L_2$. $V_B$ and $V_C$ will be increased and decreased, respectively.

*Mode 3* $(Q_x, Q_1, Q_2) = (1,0,1)$: In this mode, the input supply provides energy to phase C as well as to the inductors ($L_x, L_1$) and some of output capacitor ($C_4$) energy will be transferred to the inductor $L_1$. $V_B$ and $V_C$ will be decreased and increased, respectively.

*Mode 4* $(Q_x, Q_1, Q_2) = (1,1,1)$: In this mode, both output capacitors ($C_3, C_4$) will be discharged and the input supply provides energy to $L_X$. As a result, $V_B$ and $V_C$ will be decreased.

*Mode 5* $(Q_x, Q_1, Q_2) = (0,1,1)$: The loads is supplied via output capacitors ($C_3, C_4$). Thus output capacitors will be discharged. In this case, $L_X$ will be discharged and $V_B$ and $V_C$ will be decreased.

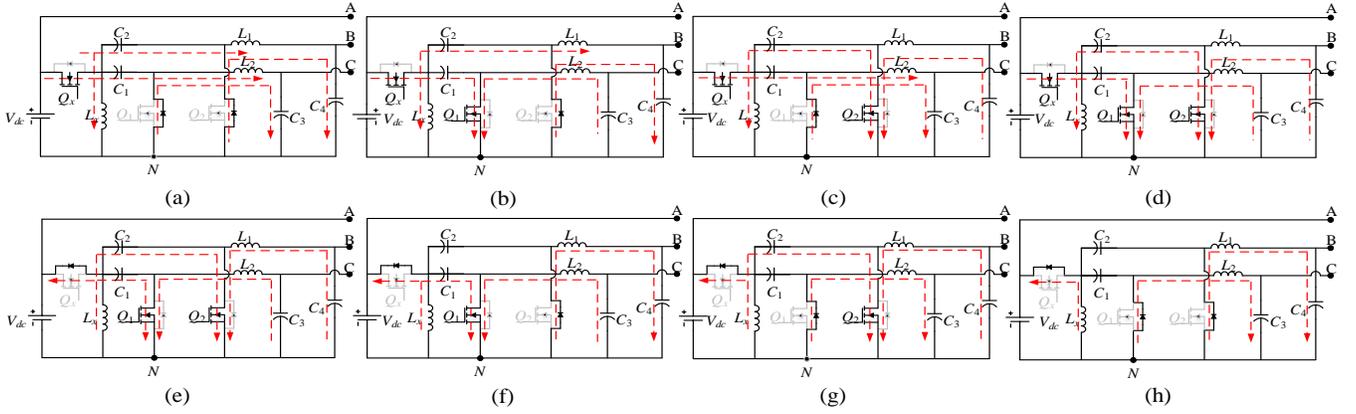

Fig. 3. Eight possible switching states of the three-switch three-phase inverter. (a) Mode 1 (Qx,Q1,Q2)=(1,0,0), (b) Mode 2 (Qx,Q1,Q2)=(1,1,0), (c) Mode 3 (Qx,Q1,Q2)=(1,0,1), (d) Mode 4 (Qx,Q1,Q2)=(1,1,1), (e) Mode 5 (Qx,Q1,Q2)=(0,1,1), (f) Mode 6 (Qx,Q1,Q2)=(0,1,0), (g) Mode 7 (Qx,Q1,Q2)=(0,0,1), (h) Mode 8 (Qx,Q1,Q2)=(0,0,0), where "1" means is the corresponding switch turned ON.

*Mode 6* $(Q_x,Q_1,Q_2)=(0,1,0)$: The energy stored in output inductor $L_1$ is used to supply the load (phase C) and some of inductor $L_1$ energy will be transferred to the output capacitor ($C_4$). In this case, ($L_X,L_1$) will be discharged and $V_B$ and $V_C$ will be decreased.

*Mode 7* $(Q_x,Q_1,Q_2)=(0,0,1)$: The energy stored in output inductor $L_2$ is used to supply the load (phase B) and some of inductor $L_2$ energy will be transferred to the output capacitor ($C_3$). In this case, ($L_X,L_2$) will be discharged and $V_B$ and $V_C$ will be decreased.

*Mode 8* $(Q_x,Q_1,Q_2)=(0,0,0)$: The energy stored in output inductors ($L_1,L_2$) are used to supply the loads (phase B and phase C) and some of output inductors energy will be transferred to the output capacitors ($C_3,C_4$). In this case, ($L_X,L_1,L_2$) will be discharged and $V_B$ and $V_C$ will be decreased.

To achieve variable output voltages commanded by reference signals there will be a need for a suitable control strategy. In this paper, the predictive controller is used to make capability of following reference signals smoothly and accurately that is explained in detail in section III.

## III. MODEL PREDICTIVE CONTROL

Considering the state-of-the-art and emerging control strategies and its trend toward better steady-state and faster transient response, the finite control set model predictive control (FCS-MPC) is applied for the proposed DC-AC converter. In FCS-MPC, the nonlinear nature of the power converters and time-domain constraints are incorporated into the control problem as a constrained optimization problem in a unified and clear manner [32]-[34]. Further, FCS-MPC provide a strong alternative to conventional PI controllers, and it can direct apply the switching states to the system. Therefore, it does not require a modulator. A disadvantage of the model predictive control is the high computational burden that increases exponentially with the length of the prediction horizon and number of switching states. Nevertheless, with the development of convex optimization techniques and computational power of digital devices, such as microprocessors, digital signal processors (DSPs), field programmable gate arrays (FPGAs), the model predictive control has become more and more popular and feasible. This has made it possible to predict the behavior of variables in standard control hardware platforms within sampling intervals of less than a few tens of $\mu$s. Actually, the most important step to design a robust controller, is a precisely modeling, that reveal the insight of the converter's dynamic behavior. The suitable mathematical model for the application of MPC is to use a state space modeling [35]-[37]. This state-space model will help to build the MPC controller. The capacitors and inductors in the proposed inverter are assumed to be non-ideal, and they have been represented by their corresponding internal resistance of the inductors and the equivalent series resistance (ESR) of the capacitors. The inverter ac-side voltages in state-space form can be expressed in terms of switching signals, capacitor voltages, inductor currents and dc-side voltage, as follows:

$$\frac{d}{dt}\begin{bmatrix}V_{BN}\\V_{CN}\\i_{Lx}\\i_{L1}\\i_{L2}\end{bmatrix}=G_{5\times 5}\begin{bmatrix}V_{BN}\\V_{CN}\\i_{Lx}\\i_{L1}\\i_{L2}\end{bmatrix}+H_{5\times 3}\begin{bmatrix}V_{C1}\\V_{C2}\\V_{dc}\end{bmatrix} \quad (11)$$

The coefficient matrices ($G,H$) can be obtained from (12), at the top of the next page, and (13) as follows:

$$H=\begin{bmatrix}0 & 0 & 0\\0 & 0 & 0\\0 & 0 & \dfrac{Q_x}{L_x}\\\dfrac{Q_x(1-Q_2)}{L_1} & 0 & \dfrac{Q_x(1-Q_2)}{L_1}\\0 & \dfrac{Q_x(1-Q_1)}{L_2} & \dfrac{Q_x(1-Q_1)}{L_2}\end{bmatrix} \quad (13)$$

An exact discretization by the zero-order hold (ZOH) method is used in this paper for digital implementation of finite control set MPC as follows:

$$\begin{bmatrix}V_{BN}(k+h)\\V_{CN}(k+h)\\i_{Lx}(k+h)\\i_{L1}(k+h)\\i_{L2}(k+h)\end{bmatrix}=\Phi_{5\times 5}\begin{bmatrix}V_{BN}(k+h-1)\\V_{CN}(k+h-1)\\i_{Lx}(k+h-1)\\i_{L1}(k+h-1)\\i_{L2}(k+h-1)\end{bmatrix}+\Gamma_{5\times 3}\begin{bmatrix}V_{C1}(k+h-1)\\V_{C2}(k+h-1)\\V_{g}(k+h-1)\end{bmatrix} \quad (14)$$

Where

$$\Phi=e^{GT_s},\ \Gamma=G^{-1}(\Phi-I_{5\times 5})H. \quad (15)$$

Common methods for model predictive control suffer from the high computational burden when solving control problems

$$G = \begin{bmatrix} -\dfrac{1}{C_4(r_{C4}+R)} & 0 & 0 & \dfrac{R}{C_4(r_{C4}+R)}(1-2Q_2) & 0 \\ 0 & -\dfrac{1}{C_3(r_{C3}+R)} & 0 & 0 & \dfrac{R}{C_3(r_{C3}+R)}(1-2Q_1) \\ 0 & 0 & -\dfrac{r_{Lx}Q_x}{L_x} + (1-Q_x)\begin{bmatrix} \dfrac{r_{C1}r_{C2}(2Q_1-1)}{r_{C1}+r_{C2}} + \\ \dfrac{(r_{C1}-r_{C2})(Q_2-Q_1)}{r_{C1}+r_{C2}} \\ +r_{Lx}(Q_1+Q_2-1) \end{bmatrix} & 0 & 0 \\ -\dfrac{R}{L_1(r_{C4}+R)} & 0 & 0 & \dfrac{r_{C4}R(2Q_2-1)}{L_1C_4(r_{C4}+R)} + \dfrac{[(r_{L1}r_{C4}+r_{L1}R)(1+2Q_xQ_2-2Q_x-2Q_2)]}{L_1C_4(r_{C4}+R)} & 0 \\ 0 & 0 & 0 & 0 & \dfrac{[(r_{L2}r_{C3}+r_{L2}R)(1+2Q_xQ_1-2Q_x-2Q_1)]}{L_2C_3(r_{C3}+R)} + \dfrac{r_{C3}R(2Q_1-1)}{L_2C_3(r_{C3}+R)} \end{bmatrix} \quad (12)$$

over long prediction horizons. In this paper, the simplified two-step prediction [38]-[39], is extended for the control of proposed inverter. Fig. 4 (a) and (b), represent the concept of prediction of control variables using two-step prediction ($h=2$) and simplified two-step prediction, respectively. Similar to one-step prediction, this method uses 8 switching states for ($k +2$) sampling instant. From a computational point of view, this makes better the solution of MPC problems and enables sampling intervals of less than a few tens of $\mu$s. The discrete-time state-space model for the two-step model predictive control can be derived as follows:

$$\begin{bmatrix} V_{BN}(k+h+2) \\ V_{CN}(k+h+2) \\ i_{Lx}(k+h+2) \\ i_{L1}(k+h+2) \\ i_{L2}(k+h+2) \end{bmatrix} = \Phi(Q_x(k),Q_1(k),Q_2(k)) \begin{bmatrix} V_{BN}(k+h+1) \\ V_{CN}(k+h+1) \\ i_{Lx}(k+h+1) \\ i_{L1}(k+h+1) \\ i_{L2}(k+h+1) \end{bmatrix} \quad (16)$$
$$+ \Gamma(Q_x(k),Q_1(k),Q_2(k)) \begin{bmatrix} V_{C1}(k+h+1) \\ V_{C2}(k+h+1) \\ V_g(k+h+1) \end{bmatrix}$$

Where $V_{BN}(k+h+1)$, $V_{CN}(k+h+1)$, and $i_L(k+h+1)$ are predicted ac-side voltages and inductor current in ($k +1$) sampling instant, respectively. These can be calculated as follows:

$$\begin{bmatrix} V_{BN}(k+h+1) \\ V_{CN}(k+h+1) \\ i_{Lx}(k+h+1) \\ i_{L1}(k+h+1) \\ i_{L2}(k+h+1) \end{bmatrix} = \Phi(Q_x(k),Q_1(k),Q_2(k)) \begin{bmatrix} V_{BN}(k+h) \\ V_{CN}(k+h) \\ i_{Lx}(k+h) \\ i_{L1}(k+h) \\ i_{L2}(k+h) \end{bmatrix} \quad (17)$$
$$+ \Gamma(Q_x(k),Q_1(k),Q_2(k)) \begin{bmatrix} V_{C1}(k+h) \\ V_{C2}(k+h) \\ V_g(k+h) \end{bmatrix}$$

$$J = \sum_{i\in\{B,C\}} \left( \sum_{j=1}^{N_p} \left(V_{iN}(k+j) - V_{Ref,iN}(k+j)\right)^2 \right) + \sum_{p\in\{x,1,2\}} \lambda_p \left(i_{Lp}(k) - \beta_p i_{Sat,p}\right)^2 \quad (18)$$

According to the model predictive control (MPC) theory, control objective is expressed as (18) cost function. The predict horizon and Lagrange multiplier are denoted in cost function as $N_P$ and $\lambda$, respectively. As shown in (18), the cost function consists of two terms. First term calculates the quadratic sum of output voltage errors. Second term penalizes the saturation in the inductors. This guarantee the inductors to work without significant inductance drop. $\beta$ is the upper bound for $I_L$. This coefficient is adjusted to regulate inductor current in an appropriate varying range. Also, the Lagrange multiplier ($\lambda$) determines the balance between the voltage control and output ripple. All possible future states of the proposed converter will be predicted. Then, the switching state which minimizes the cost function ($S_{OP}$) is selected and then applied to converter as the optimal control action for next time step. Fig. 5 illustrates a flowchart of overall required steps in implementing MPC algorithm for proposed inverter.

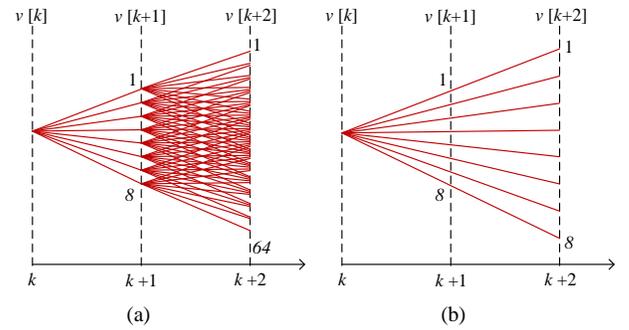

Fig. 4. Prediction of control variables using (a) two-step prediction, (b) simplified two-step prediction.

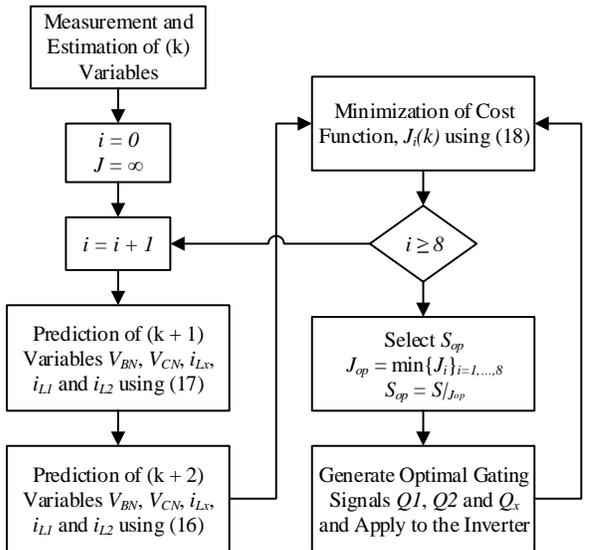

Fig. 5. Flowchart of the control strategy.

## IV. DESIGN GUIDELINES OF THE COMPONENTS

The values of the circuit components (inductors and capacitors) need to be optimized to make capability of following reference signals smoothly and accurately. In the following subsections, the derivation of components values is addressed in detail.

### A. Intermediate Capacitors

In the first step of the design, the instantaneous duty cycles for phase B and phase C are obtained as a function of the input voltage ($V_{IN}=V_{dc}$) and output voltages ($V_{out}=V_{BN}, V_{CN}$).

$$D(t) = \frac{V_{out}}{V_{IN} + V_{out}} \quad (19)$$

Hence, the instantaneous duty-cycle of third phase $D_C(t)$, can be defined as follows:

$$D_C(t) = \frac{V_{dc} + V_m \sin(\omega t - \frac{\pi}{3})}{2V_{dc} + V_m \sin(\omega t - \frac{\pi}{3})} \quad (20)$$

The worst case design occurs when the phase current is in its peak value. According to (10), the maximum value of phase current is when:

$$\omega t = \varphi + \pi \quad (21)$$

Using (21), the equation (20) can be converted to the following:

$$D_C(t)|_{i_C(t)=I_m} = \frac{V_{dc} + V_m \sin(\phi + \frac{2\pi}{3})}{2V_{dc} + V_m \sin(\phi + \frac{2\pi}{3})} \quad (22)$$

For $0 \leq \varphi \leq 90°$, the instantaneous duty-cycle will be at the highest value at $\varphi = 0$. Therefore, the maximum value of duty ratio for design is expressed as:

$$D_{\max,C} = \frac{V_{dc} + 0.866V_m}{2V_{dc} + 0.866V_m} \quad (23)$$

The maximum duty ratio for phase B can also be calculated in a similar manner, as follows:

$$D_{\max,B} = \frac{V_{dc} + V_m}{2V_{dc} + V_m} \quad (24)$$

Intermediate capacitors ($C_1$, $C_2$) are used to energy transition from the input to the output in a controlled method. These capacitors are set to achieve a compromise between constant value of voltage within one cycle and variable voltage profile in line frequency period. The voltage ripple of intermediate capacitors can be calculated as follows:

$$\Delta V_{C,\text{coup}(1,2)} = \frac{D_{\max(C,B)} i_m}{C_{\text{coup}(1,2)} f_{SW,\min}} \quad (25)$$

By admitting 5% of maximum ripple voltage in the nominal conditions of operation, the values of the intermediate capacitances are derived by

$$C_{\text{coup}(1,2)} = \frac{D_{\max(C,B)} i_m}{\Delta V_{C,\text{coup}(1,2)} f_{SW,\min}} = \frac{D_{\max(C,B)} i_m}{0.05 \times V_{dc} f_{SW,\min}} \quad (26)$$

The model predictive control has a varying switching frequency. Considering time step of discrete-time operation ($T_S$), the maximum value of switching frequency ($f_{sw,max}$) can be derived as follows:

$$f_{SW,\max} = \frac{1}{2T_S} \quad (27)$$

To meet the transient-response requirements, the minimum value of switching frequency ($f_{sw,min}$) is set as $0.2 f_{sw,max}$.

### B. Output Capacitors

The output capacitors should provide desired cutoff frequency much lower than the operating frequency. This allow a good reduction of high order harmonics. The voltage ripple of output capacitors can be calculated as follows:

$$\Delta V_{C,\text{out}(3,4)} = \frac{\Delta i_{L(2,1)}}{8C_{\text{out}(3,4)} f_{SW,\min}} \quad (28)$$

By admitting 30% of maximum ripple current on the output inductors and 10% of maximum ripple voltage on the output capacitors, the size of the output capacitors can be achieved as follows:

$$C_{\text{out}(3,4)} = \frac{0.3 I_m}{0.8(V_{dc} + V_m) f_{SW,\min}} \quad (29)$$

### C. Output Inductors

The output inductors ($L_1$, $L_2$) are set to smooth the output waveforms by the attenuation of the low frequency component. The size of these inductors can be achieved on the basis of inductors current ripple.

$$\Delta i_{L(1,2)} = \frac{V_{dc} D_{\max(B,C)}}{L_{1,2} f_{SW,\min}} = 0.3 I_m \quad (30)$$

### D. Input Inductor

By admitting 10% of maximum ripple current on the input inductor, the size of input inductor can be achieved in a similar manner, as follows:

$$L_x = \frac{V_{dc} D_{\max}}{\Delta i_{Lx} f_{SW,\min}} = \frac{V_{dc} D_{\max}}{0.1 i_{IN,\max} f_{SW,\min}} \quad (31)$$

## V. COMPARISON OF THE PROPOSED CONVERTER WITH OTHER TOPOLOGIES

In this section, the proposed inverter is compared with other three-phase inverters (standard six-switch inverter and four-switch three-phase inverter). The new converter proposes an inverter with a less number of semiconductor switches. The multilevel inverters have a high number of power semiconductor switches, and consequently, have not been considered for comparison. The comparison results are summarized in Table I, where $V_m$ is the peak of the line-to-line output voltage. According to Table I, it can be concluded that the proposed converter is one of the most promising inverters. The proposed inverter has lower number of power switches, and consequently, lower number of gate drive circuits and snubber circuits in comparison with any other universal three-phase inverter. Also, in an attempt to quantify the total rating of switches associated with each of the compared converters, the total voltage rating of the switches (TVRS) of converters are provided that can be a good criterion to compare the cost and size of converters. The new inverter achieves 15.47%, 100% and 61.3% enhancement in the DC utilization factor compared to the standard six-switch inverter with sine PWM, four-switch three-phase inverter with PWM and four-switch three-phase inverter with SVM, respectively. Unlike other inverters, the proposed inverter has pure sinusoidal voltages. Thus, there is no need for output filters. This advantage leads to the reduction of cost and size of the inverter. To achieve the same line-to-line output voltage, the proposed inverter has lower required DC-link voltage in comparison with other three-phase inverters. This advantage causes the proposed inverter be suitable for energy sources with low output voltage. Moreover, there is no dead-band between the gate drive signals of the power switches in the proposed inverter.

TABLE I
Comparison of Proposed Converter With Other Three Phase Converters.

|  | Standard six-switch inverter | Four-switch three-phase inverter | Proposed inverter |
|---|---|---|---|
| The number of used switches, gate drives | 6 | 4 | 3 |
| TVRS | $\frac{12}{\sqrt{3}}V_m$ | $8V_m$ | $7V_m$ |
| DC voltage utilization — SPWM: $\frac{1}{2}$ | | [18] $\frac{1}{2\sqrt{3}}$ | |
| DC voltage utilization — SVPWM: $\frac{1}{\sqrt{3}}$ | | [26], [28] $\frac{1}{2\sqrt{3}}$ | $\frac{1}{\sqrt{3}}$ |
| DC voltage utilization — THIPWM: $\frac{1}{\sqrt{3}}$ | | [30] $\frac{0.62}{\sqrt{3}}$ | |
| Circulation current through the DC-link capacitors | No | [18] Yes; [26], [28] Limited; [30] Limited | No |
| Dead-band between the gate drive signals | Necessary | Necessary | Not necessary |
| Output filtering components in **3 phase** | 3 LC filter | 3 LC filter | Not necessary |
| Complexity | Simple | Average | Average |
| Power capacity | High | Medium | Medium |
| Properties | Accepted as standard topology | -Capacitor voltage balancing issues. - Access to dc-bus mid-point is Required. | Large AC pulsing components |

Nevertheless, the proposed inverter has more passive components compare to standard six-switch inverter and four-switch three-phase inverter. Thus, a direct efficiency calculation clearly leads one to two percent less efficiency compared to these converters. Also, it does increase the cost and size of proposed converter. However, it is noteworthy that in conventional six-switch inverters, the associated cost and size of extra heat sinks and gate drives for extra switches is also significant. In general, the proposed inverter is very competitive in low and medium power range.

## VI. EXPERIMENTAL RESULTS

In order to validate the performance of the proposed converter, a prototype was built as a proof-of-concept. The circuit parameters and manufacturer numbers of the components are given in Table II. To select the model predictive control parameters in the hardware implementation, the influences of the MPC parameters ($N_P$ and $\lambda$) on the total harmonic distortion are studied in simulation environment. The predict horizon ($N_P$) and Lagrange multiplier ($\lambda$) are set differently and their impacts are analyzed. Fig. 6 show the percentages of the total harmonic distortion when $T_S = 25\mu s$ and $V_m = 100^V$. From Fig. 6, it can be seen that the optimal parameter settings for the MPVC method are $1 \leq N_P \leq 2$ and $0 < \lambda \leq 0.10$. Hence, for better total harmonic distortion, small $N_P$ and $\lambda$ can be selected for the MPC method. Then considering the computational burden and to achieve a compromise between practicability and high quality, $N_P$ and $\lambda$ are set as 2 and 0.10 based on the analysis in Fig. 6.

The experimental hardware platform consists of the power circuit of proposed converter and a control board (An STM32F429IGT6 microcontroller based on ARM Cortex-M4 core). IRF 740 and TLP250 are employed as the power MOSFETs and drivers, respectively. The control board has a maximum operating frequency of 180 MHz and a maximum ADC sampling rate of 7.2 MSPS. The inverter feeds a grounded star-connected resistive load that parameters are summarized in Table II.

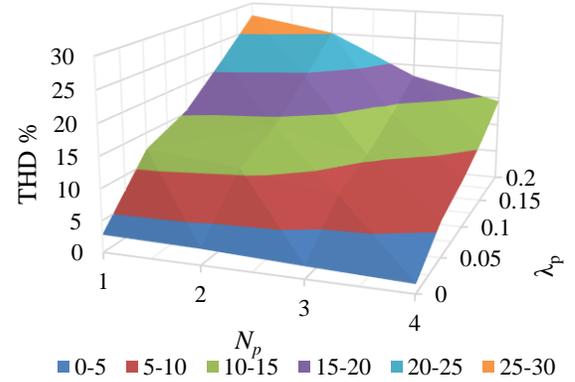

Fig. 6. Influence of the FCS model predictive control parameters (NP and λ) on the total harmonic distortion when TS = 25μs and Vm = 100V.

TABLE II
The Circuit Parameters and Manufacturer Numbers of The Converter's Components

| Symbol | Quantity or Device | Parameter |
|---|---|---|
| $C_1, C_2$ | Intermediate capacitors | Two paralleled $15^{\mu F}$ |
| $C_3, C_4$ | Output capacitors | $1^{\mu F}$ |
| $\lambda_p$ | Weighting factor | 9.9 |
| $L_x$ | Internal inductor | $3.6^{mH}$ |
| $L_1, L_2$ | Output inductors | $1.2^{mH}$ |
| $V_m$ | Peak output line voltage | $100^V$ |
| $f$ | Output frequency | $50^{Hz}$ |
| $R_L$ | Resistance load | $50\Omega$ |

The required current measurements are made by ACS712, and sent to the control board. The response time of current transducer ACS712 is 5 μs. For the inductors, the tolerable maximum value of $I_L$ is $I_{L,\max} = 0.5\ I_{sat}$ in the experiment for the MPC controller, i.e., the constraint coefficient is $\beta = 0.5$. The saturation current $I_{sat}$ of inductors $L_{1,2}$, $L_x$ are 14.7$^A$ and 17.7$^A$, respectively. The inductor current $I_L$ waveform is actually the voltage signal $V_{IL}$ from the current transducer, and the relationship between $I_L$ and $V_{IL}$ is described by a linear equation as follows:

$$I_L = 5.33\ V_{IL} - 13.33 \qquad (32)$$

Therefore, maximum values of inductor currents $I_{L,\max}$ can be calculated based on the maximum value of $V_{IL}$. These guarantees that the inductors are away from saturation.

As discussed in Section II, the basic idea behind control strategy is to produce a set of dc-biased sinusoidal voltages (phase B and phase C) with a phase difference of 60° and connect the phase A to the input dc supply to maintain balanced line-to-line voltages, which can be seen from Fig. 7. Also, it is shown that the output voltages are well regulated without any filtering requirements. The experimental waveforms of the output capacitor voltages (voltages of phase B and phase C) with respect to the supply common point are presented in Fig. 8. The experimental key waveforms of the inverter (voltage across the intermediate capacitor $C_1$ and current through the output inductor $L_1$) are shown in Figs. 9–10. In order to observe transient response of proposed inverter, it is assumed that at $t=30$, inverter has experienced a step change in the frequency and load voltage reference. The dynamic response of the proposed inverter when feeding an inductive-resistive load (R-L), is depicted in Fig. 11.

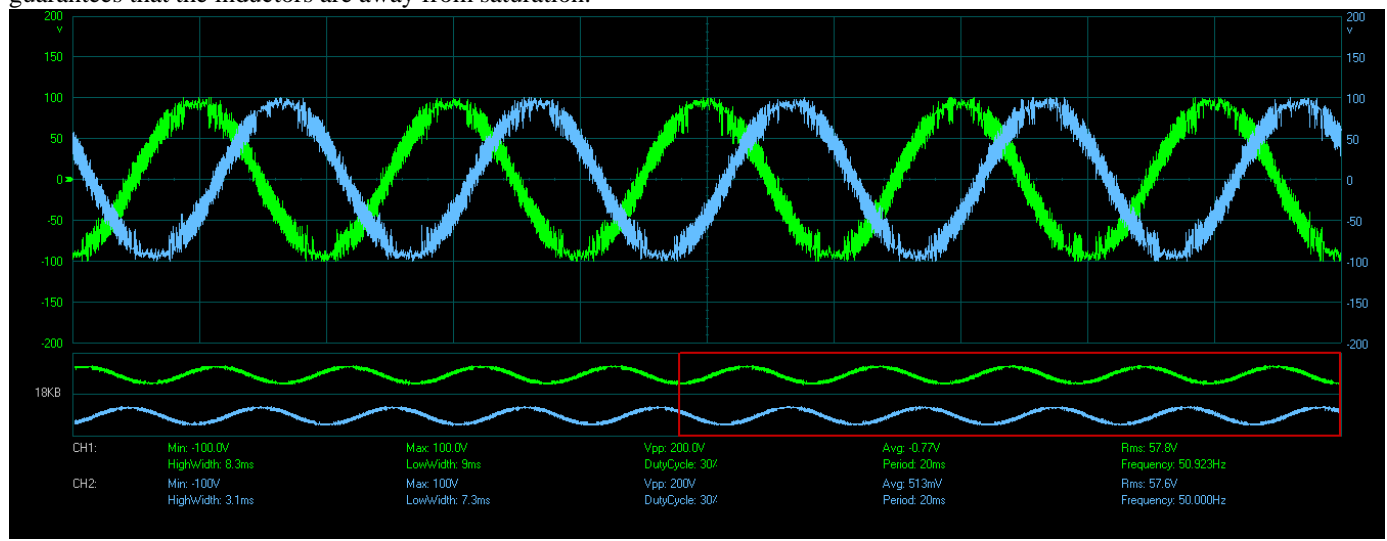

(a)

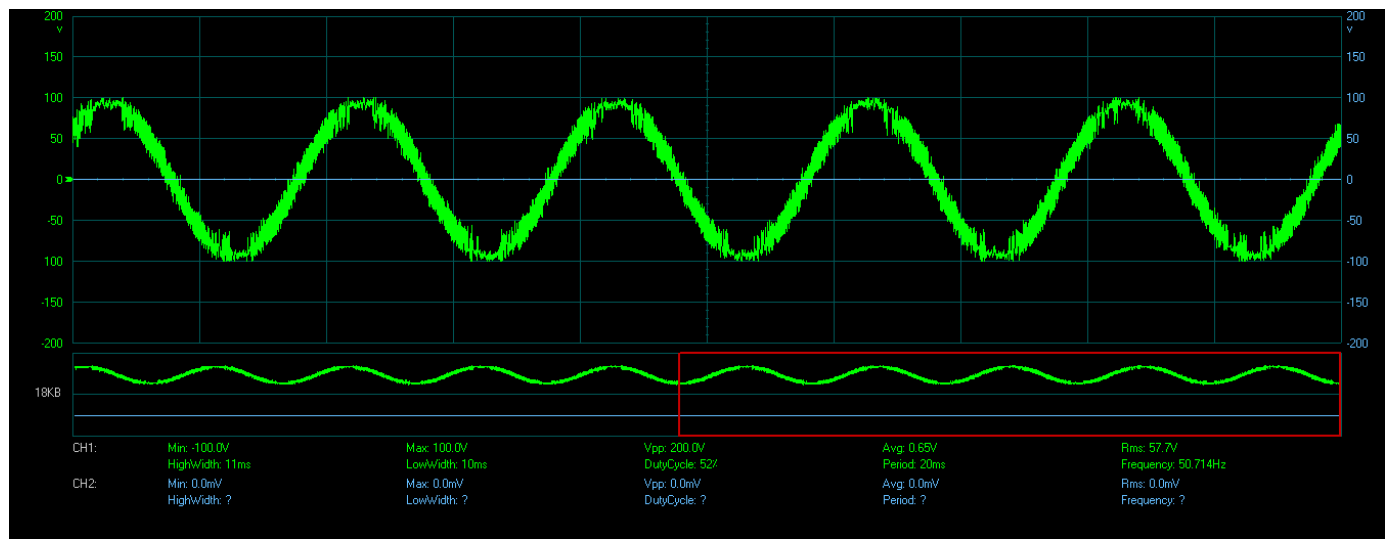

(b)

Fig. 7. Experimental waveforms of the output line voltages across the load. (a) Phase A and Phase B. (b) Phase C. (50 V/div, 10 ms/div).

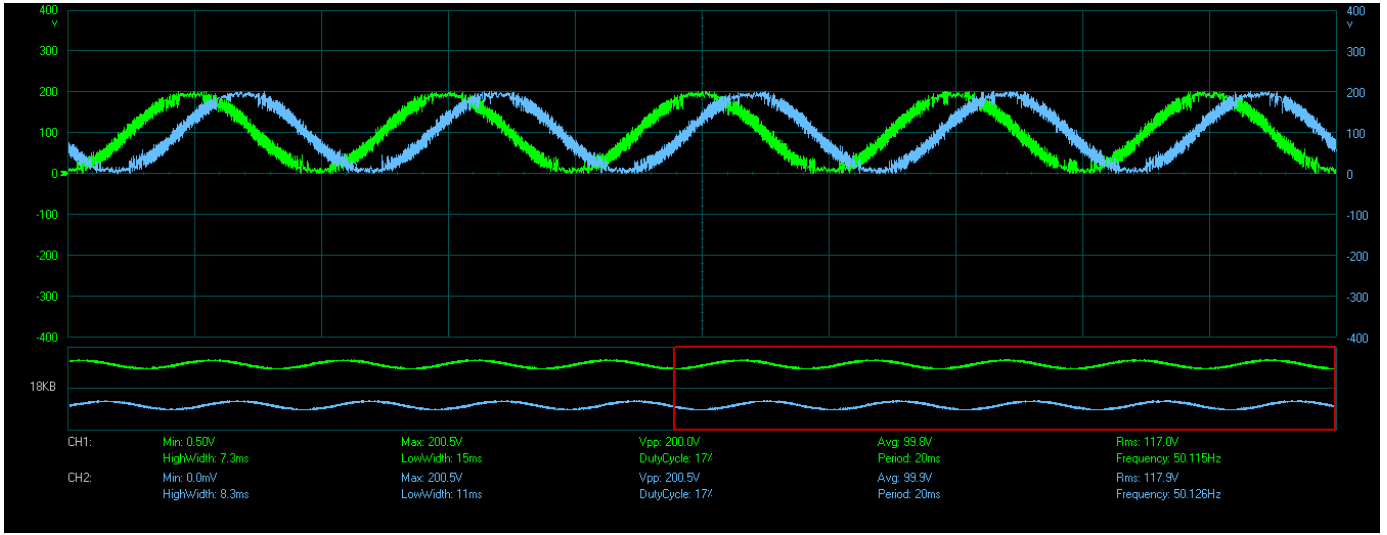
Fig. 8. Experimental waveforms of the output capacitor voltages. (100 V/div, 10 ms/div).

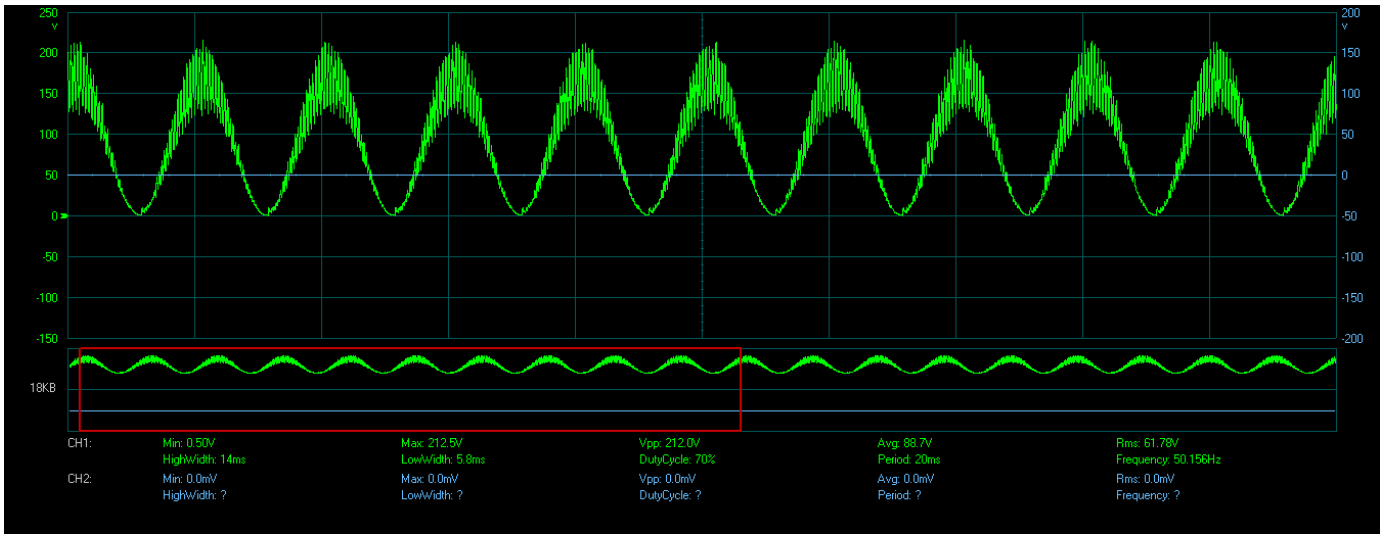
Fig. 9. Experimental waveform of the voltage across the intermediate capacitor C1. (50 V/div, 20 ms/div).

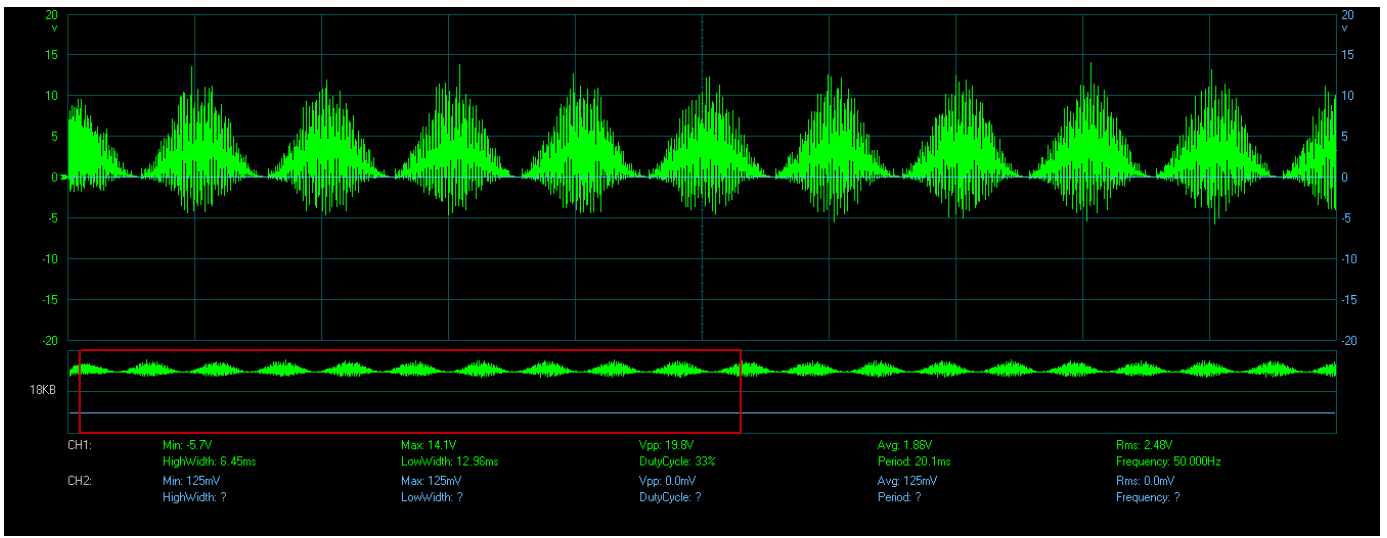
Fig. 10. Experimental waveform of the current through the output inductor L1. (5 A/div, 20 ms/div).

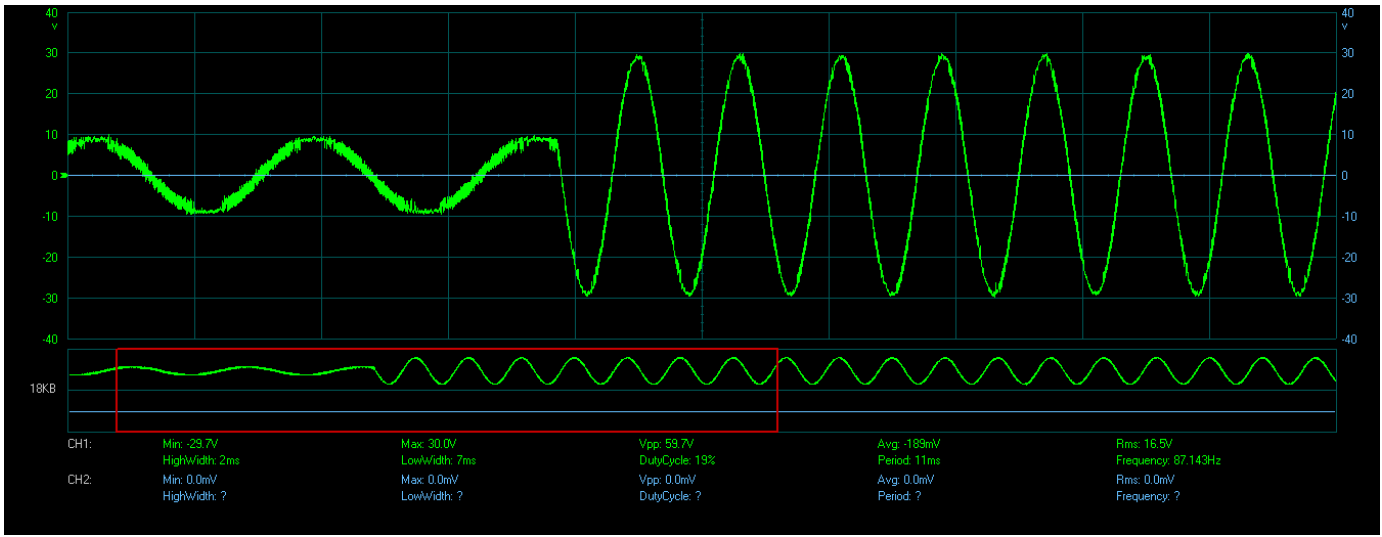

Fig. 11. Dynamic response of the proposed inverter for a step change of frequency and load voltage reference (Phase A). (10 V/div, 10 ms/div).

In experimental results, the inverter voltages effectively tracked their references even with step change of frequency and load voltage reference. This show clearly that the new inverter topology works as expected using only three power switches. Also, this topology can be operated at high frequencies with a good output spectrum.

## VII. Conclusion

A cost-effective single-stage voltage source DC–AC converter with a less number of semiconductor switches and respective control strategy have been proposed. In this paper, a simplified two-step finite control set model predictive control is extended for the control of proposed inverter. The continuous- and discrete-time modeling of the proposed inverter is presented for digital implementation. It is important to note that no delay compensation is needed with the proposed simplified two-step prediction. The proposed inverter improves the utilization of the input dc supply by a factor of 1.15 and 2 compared to the standard six-switch inverter and four-switch three-phase inverter, respectively. The output line-to-line voltages of proposed inverter are pure sinusoidal waves and total harmonic distortion is minimized. Also, there is no circulation of one-phase current through the split dc-link capacitors. In addition, dead-band between the gate drive signals of the two interlocked semiconductor devices will not be considered in proposed inverter. Experimental verification has presented promising results and it is shown that the proposed inverter has good potential for renewable energy sources with low output voltage.

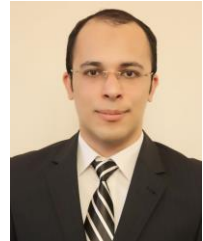

**Masoud Farhadi** received the B.Sc. degree in electrical engineering with honors and M.Sc. degree in power engineering (power electronics and systems) with honors from the Department of Electrical Engineering, University of Tabriz, Iran, in 2013 and 2016, respectively

His current research interests include analysis and control of power electronic converters, reliability of power electronic systems, and renewable energy conversion systems.

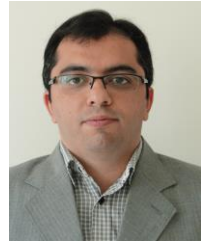

**Mehdi Abapour** (S'08–M'12) received his M.Sc. and Ph.D. degrees in electrical engineering from University of Tabriz and Tarbiat Modares University, in 2007 and 2013, respectively. He is currently an Assistant Professor with the Department of Electrical and Computer Engineering, University of Tabriz, Iran.

His research interests include stochastic power system analysis, reliability of power electronic systems and power systems, power system dynamics, FACTS devices and fault current limiters.